\documentclass[conference]{IEEEtran}
\IEEEoverridecommandlockouts
\usepackage{stmaryrd}
\usepackage{lipsum}  
\usepackage{amsmath,amsfonts,amssymb,amsthm}
\usepackage[ruled,vlined]{algorithm2e}
\usepackage{array}
\usepackage[caption=false,font=normalsize,labelfont=sf,textfont=sf]{subfig}
\usepackage{textcomp}
\usepackage{stfloats}
\usepackage{url}
\usepackage{verbatim}
\usepackage{graphicx}
\usepackage{cite}
\usepackage{xcolor}
\usepackage[nolist,nohyperlinks]{acronym} 
\usepackage{amsmath,amssymb,amsfonts}
\usepackage{algorithmic}
\usepackage{graphicx}
\usepackage{textcomp}
\usepackage{caption}
\usepackage{subcaption} 
\def\BibTeX{{\rm B\kern-.05em{\sc i\kern-.025em b}\kern-.08em
    T\kern-.1667em\lower.7ex\hbox{E}\kern-.125emX}}
\newcommand{\bb}[1]{\mathbb{#1}}
\newcommand{\ten}[1]{\boldsymbol{\mathcal #1}}
\newcommand{\ma}[1]{\boldsymbol{#1}}
    \renewcommand{\baselinestretch}{0.94}
\begin{document}
\begin{acronym}
	\acro{IRS}{Intelligent reflecting surface}
	\acro{RIS}{reconfigurable intelligent surface}
	\acro{irs}{intelligent reflecting surface}
	\acro{PARAFAC}{parallel factor}
	\acro{TALS}{trilinear alternating least squares}
	\acro{BALS}{bilinear alternating least squares}
	\acro{DF}{decode-and-forward}
	\acro{AF}{amplify-and-forward}
	\acro{CE}{channel estimation}
	\acro{RF}{radio-frequency}
	\acro{THz}{Terahertz communication}
	\acro{EVD}{eigenvalue decomposition}
	\acro{CRB}{Cramér-Rao lower bound}
	\acro{CSI}{channel state information}
	\acro{BS}{base station}
	\acro{MIMO}{multiple-input multiple-output}
	\acro{NMSE}{normalized mean squared error}
	\acro{2G}{Second Generation}
	\acro{3G}{3$^\text{rd}$~Generation}
	\acro{3GPP}{3$^\text{rd}$~Generation Partnership Project}
	\acro{4G}{4$^\text{th}$~Generation}
	\acro{5G}{5$^\text{th}$~Generation}
	\acro{6G}{6$^\text{th}$~generation}
	\acro{E-TALS}{\textit{enhanced} TALS}
	\acro{UT}{user terminal}
	\acro{UTs}{users terminal}
	\acro{LS}{least squares}
	\acro{KRF}{Khatri-Rao factorization}
	\acro{KF}{Kronecker factorization}
	\acro{MU-MIMO}{multi-user multiple-input multiple-output}
	\acro{MU-MISO}{multi-user multiple-input single-output}
	\acro{MU}{multi-user}
	\acro{SER}{symbol error rate}
	\acro{SNR}{signal-to-noise ratio}
	\acro{SVD}{singular value decomposition}
\end{acronym}

\title{Semi-Blind  Channel  Estimation for Beyond Diagonal RIS\\
}

% \author{\IEEEauthorblockN{1\textsuperscript{st} Gilderlan Tavares de Araújo}
% \IEEEauthorblockA{
% \textit{Federal Institute of Ceara (IFCE)}\\
% Canindé-CE, Brazil \\
% gilderlan@gtel.ufc.br}
% \and
% \IEEEauthorblockN{2\textsuperscript{nd} André Lima Férrer de Almeida}
% \IEEEauthorblockA{\textit{Teleinformatic Department} \\
% \textit{and Wireless Telecommunication Group (GTEL)}\\
% \textit{Federal University of Ceara (UFC)}\\
% Fortaleza-CE, Brazil \\
% andre@gtel.ufc.br}
% % \and
% % \IEEEauthorblockN{3\textsuperscript{rd} Given Name Surname}
% % \IEEEauthorblockA{\textit{dept. name of organization (of Aff.)} \\
% % \textit{name of organization (of Aff.)}\\
% % City, Country \\
% % email address or ORCID}
% % \and
% % \IEEEauthorblockN{4\textsuperscript{th} Given Name Surname}
% % \IEEEauthorblockA{\textit{dept. name of organization (of Aff.)} \\
% % \textit{name of organization (of Aff.)}\\
% % City, Country \\
% % email address or ORCID}
% % \and
% % \IEEEauthorblockN{5\textsuperscript{th} Given Name Surname}
% % \IEEEauthorblockA{\textit{dept. name of organization (of Aff.)} \\
% % \textit{name of organization (of Aff.)}\\
% % City, Country \\
% % email address or ORCID}
% % \and
% % \IEEEauthorblockN{6\textsuperscript{th} Given Name Surname}
% % \IEEEauthorblockA{\textit{dept. name of organization (of Aff.)} \\
% % \textit{name of organization (of Aff.)}\\
% % City, Country \\
% % email address or ORCID}
% }

\author{Gilderlan Tavares de Araújo$^1$ and André L. F. de Almeida$^2$\\
\textit{Federal Institute of Ceara}$^1$, Caninde, Brazil\\
\textit{Federal University of Ceara}$^2$, Fortaleza, Brazil\\
\{gilderlan,andre\}@gtel.ufc.br
        % <-this % stops a space
\thanks{The authors acknowledge the partial support of Fundação Cearense de Apoio ao Desenvolvimento Científico e Tecnológico (FUNCAP) under grants FC3-00198-00056.01.00/22 and  ITR-0214-00041.01.00/23, and the National Institute of Science and Technology (INCT-Signals) sponsored by Brazil's National Council for Scientific and Technological Development (CNPq) under grant 406517/2022-3. This work is also partially supported by CNPq under grants 312491/2020-4 and 443272/2023-9.}% <-this % stops a space
%\thanks{Manuscript received April 19, 2021; revised August 16, 2021.}
}

\maketitle

\begin{abstract}
The channel estimation problem has been widely discussed in traditional \acl{RIS} assisted \acl{MIMO}. However, solutions for channel estimation adapted to beyond diagonal RIS need further study, and few recent works have been proposed to tackle this problem. Moreover, methods that avoid or minimize the use of pilot sequences are of interest. This work formulates a data-driven (semi-blind) joint channel and symbol estimation algorithm for beyond diagonal RIS that avoids a prior pilot-assisted stage while providing decoupled estimates of the involved communication channels. The proposed receiver builds upon a PARATUCK tensor model for the received signal, from which a trilinear alternating estimation scheme is derived. Preliminary numerical results demonstrate the proposed method's performance for selected system setups. The symbol error rate performance is also compared with that of a linear receiver operating with perfect knowledge of the cascaded channel. \end{abstract}


\section{Introduction}
\IEEEPARstart{R}{econfigurable} intelligent surfaces (RISs) were introduced a few years ago as a potential technology for beyond 5G or 6G wireless systems. However, this technology involves several challenges, including channel estimation, passive and active beamforming optimization, and multi-user interference \cite{zheng_survey}. Concerning the channel estimation problem, most of the work in this area has focused on the traditional RIS architecture, where the phase shift matrix is diagonal, which means that physical coupling among RIS elements is not considered. A new RIS architecture with a non-diagonal RIS phase shift matrix was recently proposed in \cite{Clerckx_TWC_APR_2023, B_Clerckc_CE_TSP_24}. The so-called beyond diagonal RIS (BD-RIS) adds degrees of freedom for system optimization and provides performance gains compared to diagonal RIS.  The non-diagonal structure of the phase shift matrix also arises in non-reciprocal connections of RIS elements, enabling an incident signal to be reflected from another element \cite{Clerckx_TVT_JUN_2022}. The \cite{Shen_2022} proposed a modeling and architecture design based on scattering network analysis. More recently, the work \cite{matteo_2024} presented a RIS modeling framework based on multiport network analysis., while optimization algorithms for BD-RIS were discussed in \cite{Clerckx_TWC_FEB_2024, B.Clerckx_TWC_EA_2024}.

The promised performance enhancements offered by BD-RIS (compared to traditional/single-connected RIS) highly depend on the accuracy of the channel state information (CSI). The BD-RIS's lack of signal processing capabilities, combined with the connection structure of more complex elements, makes channel estimation even more challenging than single-connected RIS. Few solutions for channel estimation for BD-RIS have been proposed in the literature. The work \cite{B_Clerckc_CE_TSP_24} proposed a least squares (LS) method to estimate the cascaded channel, which requires a large training overhead. In \cite{Sokal_BD_RIS_2024}, closed-form and iterative algorithms for BD-RIS channel estimation were proposed by capitalizing on tensor decomposition. While \cite{B_Clerckc_CE_TSP_24} only estimates the cascaded channel, the tensor-based methods of \cite{Sokal_BD_RIS_2024} provide individual estimates of the two involved channel matrices while reducing the training overhead. However, these methods resort to pilot sequences, implying that CSI acquisition and data detection are carried out in two consecutive stages.

This paper investigates a data-driven (semi-blind) joint channel and symbol estimation for BD-RIS while providing decoupled estimates of the involved communication channels. Assuming a two-time scale transmission protocol along with an adapted scattering matrix design, we propose a semi-blind receiver that provides data-aided channel estimation by resorting to a tensor modeling of the received signals according to the PARATUCK decomposition \cite{Ximenes_2014,Gil_TSP}. The proposed PARaatuck rEceiver (PARE) is based on an alternating linear estimation scheme \cite{Pierre_2009} that ensures unique estimates (up to scaling ambiguities) of the channels and symbol matrices.

Tensor decompositions have been successfully applied to solve joint channel estimation and data detection problems in several wireless communication systems \cite{Almeida_Elsevier_2007,Almeida2008,Favier2014,Ximenes_2015,Chen2021}, and, more recently, to solve the individual channel estimation for reconfigurable surfaces \cite{Araujo2020SAM,Gil_JTSP}. The algorithm derived in this paper avoids a prior pilot-assisted stage by intertwining symbol and channel estimation. Numerical results evaluate the performance of the proposed method for some selected system setups. Symbol estimation performance is assessed in terms of the bit error rate and compared with that of a linear receiver under perfect knowledge of the cascaded channel. 

%\vspace{-2ex}
%\noindent \textit{Notation and properties}: Matrices are represented with boldface capital letters ($\ma{A})$, and vectors are denoted by boldface lowercase letters ($\ma{a})$. Tensors are symbolized by calligraphic letters $(\ten{A})$. Transpose and pseudo-inverse of a matrix $\ma{A}$ are denoted as $\ma{A}^{\textrm{T}}$ and $\ma{A}^\dagger$, respectively.  $D_i(\ma{A})$ is a diagonal matrix holding the $i$th row of $\ma{A}$ on its main diagonal. The operator $\diamond$ and $\otimes$ denote the Khatri Rao and Kronecker products, respectively. $\ma{I}_N$ denotes the $N \times N$ identity matrix. The operator $\textrm{vec}(\cdot)$ vectorizes an $I \times J$ matrix argument, while $\textrm{unvec}_{I \times J}(\cdot)$ does the opposite operation. Moreover, $\textrm{vecd}(.)$ forms a vector out of the diagonal of its matrix argument. ¨The $n$-mode unfolding of  a tensor $\ten{Y} \in \bb{C}^{I \times J \ldots \times N}$ is denoted by $[\ten{Y}]_{(n)}$. Moreover, $\ma{A}_{i.}$ and $\ma{A}_{.j}$ denotes the $i$-th row and $j$-th column of the matrix $\ma{A}$, respectively. In this paper, we make use of the following identities:
% \begin{equation}
% (\ma{A} \diamond \ma{B})^{\textrm{H}}(\ma{C} \diamond \ma{D}) = (\ma{A}^{\textrm{H}}\ma{C}) \odot  (\ma{B}^{\textrm{H}}\ma{D}).
% \label{Eq:Propertie Hadmard x Khatri}
% \end{equation}
\noindent \textit{Notation and properties}: In this paper, we make use of the following identities:
\begin{equation}
\textrm{vec}(\ma{ABC}) = (\ma{C}^{\textrm{T}} \otimes \ma{A})\textrm{vec}(\ma{B}).
\label{Eq:Propertie Vec General}
\end{equation}
\begin{equation}
\textrm{diag}(\ma{a})\ma{b} = \textrm{diag}(\ma{b})\ma{a}.
\label{Eq:Propertie diag(a)b}
\end{equation}
If $\ma{B}$ is a diagonal matrix, we have: 
\begin{equation}
\textrm{vec}(\ma{ABC}) = (\ma{C}^{\textrm{T}} \diamond \ma{A})\textrm{vecd}(\ma{B}).
\label{Eq:Propertie Vec restrict}
\end{equation}

\subsection{PARATUCK decomposition}
\label{Overview tensor}
The PARATUCK decomposition \cite{Harshman_1996} is a hybrid tensor decomposition combining the Tucker \cite{Tucker1966} and the PARAFAC decompositions. It enjoys unique properties under mild conditions while offering a more flexible structure compared to PARAFAC. Its scalar and slice representations are given as 
\begin{equation}
x_{i,j,k} = \sum_{r_1 = 1}^{R_1}\sum_{r_2 = 2}^{R_2}a_{i,r_1}b_{j,r_2}\omega_{r_1,r_2}c_{k,r_1}^{A}c_{k,r_2}^{B},
\label{PT2:scalarNotation}
\end{equation}
and 
\begin{equation}
\ma{X}_k = \ma{A}D_k(\ma{C}^A)\ma{\Omega}D_k(\ma{C}^B)\ma{B}^{\textrm{T}},
\label{PT2:SliceNotation}
\end{equation}
respectively, where $a_{i,r_1}$, $b_{j,r_2}$, $\omega_{r_1,r_2}$, $c_{k,r_1}^{A}$ and $c_{k,r_2}^{B}$ are the elements of the matrices $\ma{A} \in \bb{C}^{I \times R_1}$, $\ma{B} \in \times \bb{C}^{J \times R_2}$,  $\ma{\Omega} \in \bb{C} ^{R_1 \times R_2}$, $\ma{C}^{A} \in \bb{C}^{K \times R_1}$ and $\ma{C}^{B} \in \bb{C}^{K \times R_2}$, respectively. $\ma{A}$ and $\ma{B}$ are referred to as the \emph{factors matrices}, $\ma{C}^{A}$ and $\ma{C}^{B}$ are the \emph{interactions matrices}, while $\ma{\Omega}$ is the \emph{core matrix}, whose $(r_1,r_2)$-th entry defines the level of interaction between the $r_1$-th column of $\ma{A}$ and the $r_2$-th column of $\ma{B}$. 

\section{System model and assumptions}
\label{SEC:System_Model}

\begin{figure}[!t]
    \centering
    \begin{minipage}[b]{0.45\textwidth}
        \centering
        \includegraphics[width=\textwidth]{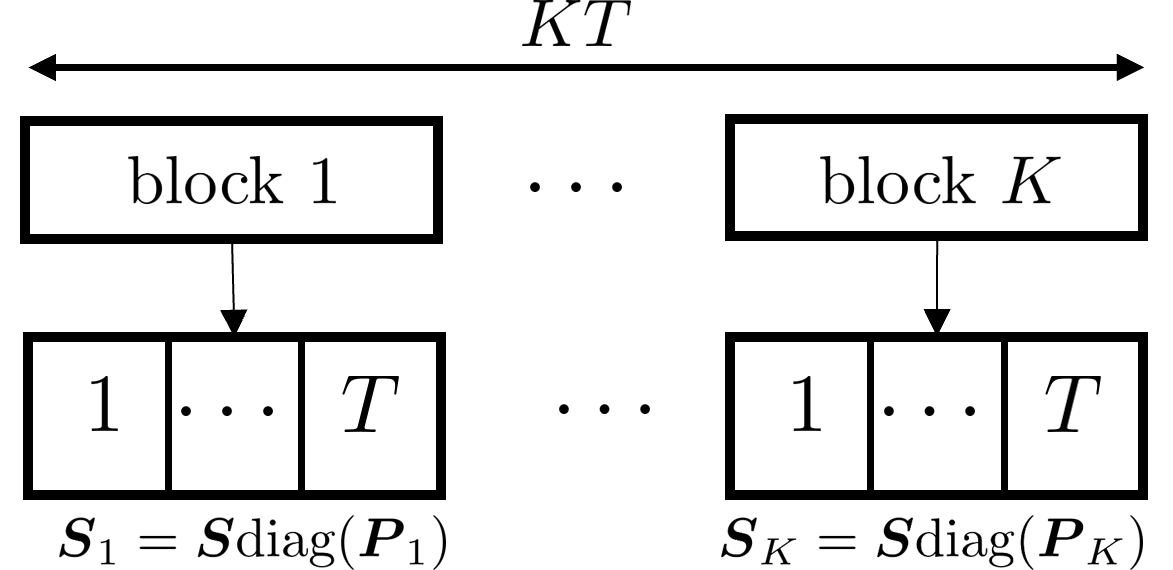}
        \caption{Transmission protocol.}
        \label{fig:time}
    \end{minipage}
    \hfill
   % \hspace{2cm}
   %  \begin{minipage}[b]{0.5\textwidth}
   %      \centering
   %      \includegraphics[width=\textwidth]{Asilomar_2024/fig/PT2_V2.png}
   %      \caption{\small{PARATUCK decomposition of the received signal.}}
   %      \label{fig:PARATUCK2}
   %  \end{minipage}
   % % \caption{Legenda geral}
   %  \label{fig:geral}
\end{figure}

Consider a %narrowband
single-user RIS-assisted MIMO system where the transmitter and receiver have $M_T$ and $M_R$ antennas, respectively. The coherence time is divided into an uplink time $T_u$ and a download time $T_d$, and the semi-blind channel and symbol estimation occur on the uplink side. The RIS has $N$ elements, but the direct link is unavailable. We assume a group-connected RIS where $Q$ groups are composed of $N_q = \frac{N}{Q}$ elements each, so all elements within each group are connected. As a result, the scattering matrix has a non-diagonal structure {\color{black} and the received signal is given as \cite{Clerckx_CAMSAP_2023}
\begin{equation}
    \ma{Y}_{k,t} = \sum\limits_{q=1}^Q\ma{H}^{(q)}\ma{S}_{k}^{(q)}\ma{G}^{(q)}\ma{x}_t + \ma{z}_{t}.
    \label{Eq: RX signal reference}
 \end{equation}
Let us consider that the} transmission occurs during $K$ time blocks of length $T$ %, each composed of $T$ time slots 
as shown in Figure \ref{fig:time}. Each block transmits different sets of symbols, while the scattering matrix remains constant within a block and varies from block to block. {\color{black}Applying a block-wise Khatri-Rao coding at the transmission,} the signal received at block $k$ {\color{black} and time slot $t$} is given by
\begin{equation}
    \ma{Y}_{k} = \sum\limits_{q=1}^Q\ma{H}^{(q)}\ma{S}_{k}^{(q)}\ma{G}^{(q)}\text{diag}(\ma{w}_{k})\ma{x}_{k,t} + \ma{z}_{k,t},
    \label{Eq: Block DB_RIS_PT2}
 \end{equation}
 {\color{black}such that, after $T$ time slots, we have
 \begin{equation}
    \ma{Y}_{k} = \sum\limits_{q=1}^Q\ma{H}^{(q)}\ma{S}_{k}^{(q)}\ma{G}^{(q)}\text{diag}(\ma{w}_{k})\ma{X}^{\text{T}} + \ma{Z}_{k},
    \label{Eq: Block DB_RIS_PT2}
 \end{equation}
 }where $\ma{X} = \left[\ma{x}_1 \; \dots \; \ma{x}_T\right]^{\text{T}} \in \bb{C}^{T \times M_T}$ denotes the data symbol matrix, $\ma{H}^{(q)}$ and  $\ma{G}^{(q)}$ are the RIS-BS and UT-RIS channels respectively, $\ma{S}_k$ is the nondiagonal scattering matrix, and $\ma{Z}_{k}$ is the additive white Gaussian noise (AWGN) term.

{\color{black}
\noindent \subsubsection*{Scattering matrix design}

Starting from the fundamental signal model in (\ref{Eq: Block DB_RIS_PT2}), we propose to design the BD-RIS training matrix as a product of a fixed scattering matrix $\ma{S}_0$ and a block-dependent diagonal rotation, such that  $\ma{S}_k^{(q)} = \ma{S}_0\text{diag}(\ma{\overline{p}}_{k}^{(q)})$ where $\ma{\overline{p}}_{k,q}$ is the vector containing the phase rotation angles associated with the $q$th group and $k$th block subject to $(\ma{S}_{k}^{(q)})^{\text{H}}\ma{S}_{k}^{(q)} = \ma{S}_{k}^{(q)}(\ma{S}_{k}^{(q)})^{\text{H}} = \ma{I}_{N_q}.  $
% \begin{equation}
% (\ma{S}_{k}^{(q)})^{\text{H}}\ma{S}_{k}^{(q)} = \ma{S}_{k}^{(q)}(\ma{S}_{k}^{(q)})^{\text{H}} = \ma{I}_{N_q}.  
% \label{EQ: identity condition for S_k}
% \end{equation}
}

\section{Semi-blind PARatuck rEceiver (PARE)}
% Starting from the fundamental signal model in (\ref{Eq: Block DB_RIS_PT2}), we propose to design the BD-RIS training matrix as a product of a fixed scattering matrix $\ma{S}_0$ and a block-dependent diagonal rotation, such that  $\ma{S}_k^{(q)} = \ma{S}_0\text{diag}(\ma{\overline{p}}_{k}^{(q)})$ where $\ma{\overline{p}}_{k,q}$ is the vector containing the phase rotation angles associated with the $q$th group and $k$th block. 
Assuming the noiseless case for notation convenience, the received signal can be expressed as 
\begin{equation}
\begin{aligned}
  \ma{Y}_{k} & = \left(\sum_{q = 1}^{Q}\ma{H}^{(q)}\ma{S}_{0}^{(q)}\text{diag}(\ma{\overline{p}}_{k}^{(q)})\ma{G}^{(q)}\text{diag}(\ma{w}_{k})\ma{X}^{\text{T}}\right) \\
  & = \ma{H}\ma{S}\text{diag}(\ma{p}_{k})\ma{G}\text{diag}(\ma{w}_{k})\ma{X}^{\text{T}},
\label{Eq:Rx siganl at (k,t)}   
\end{aligned}
\end{equation}
where $\ma{H}  \doteq \left[\ma{H}_1 \; \dots \; \ma{H}_Q\right] \in \ma{C}^{M_R \times N}$, $\ma{G}  \doteq \left[\ma{G}_{1}^{\text{T}} \; \dots \; \ma{G}_{Q}^{\text{T}}\right]^{\text{T}} \in \ma{C}^{N \times M_T}$ are the Rx-RIS and TX-RIS channels, respectively, while $\ma{p}_k\doteq[\ma{p}^{(1)\text{T}}_k, \ldots, \ma{p}^{(Q)\text{T}}_k]^{\text{T}} \in \bb{C}^{N \times 1}$, and $\ma{S} \doteq \text{bdiag}(\ma{S}_{0}^{(1)}, \ldots \ma{S}_{0}^{(Q)}) \in \bb{C}^{N \times N}$, with $\ma{S}_{0}^{(q)} \ma{S}_{0}^{(q)\text{H}}= \ma{I}_{N_q}$. Let us define $\ma{P}=[\ma{p}_1, \ldots, \ma{p}_K]^{\text{T}} \in \bb{C}^{K \times N}$ and $\ma{W}\doteq [\ma{w}_1, \ldots, \ma{w}_K]^{\text{T}} \in \bb{C}^{K \times M_T}$ as the BD-RIS phase rotation matrix and the transmitter side coding matrix, respectively. Additionally, let $\overline{\ma{H}} = \ma{H}\ma{S} \in \bb{C}^{M_R \times N}$ be the effective BS-RIS channel matrix. With these definitions, we can rewrite the 
noiseless received signal as the $k$-th block as
\begin{equation}
    \ma{Y}_{k} = \overline{\ma{H}}D_k(\ma{P})\ma{G}D_k(\ma{W})\ma{X}^{\text{T}} \in \bb{C}^{M_R \times T}, \,\, k=1, \ldots, K.
    \label{Eq:RX signal at k}
\end{equation}
We can view the received signal matrices $\{\ma{Y}_{1}, \ldots, \ma{Y}_{K}\}$ as frontal slices of a {\color{black} thrid-order} tensor $\ten{Y} \in \bb{C}^{M_R \times T  \times K}$. % of which are number $M_R$ of receive antennas, block length $T$, and number $K$ of data blocks. 
In fact, the signal model (\ref{Eq:RX signal at k}) corresponds to a PARATUCK tensor model \cite{Andre_EURASIP__CE_2014,Gil_TSP}. To summarize, comparing equations  (\ref{PT2:SliceNotation}) and (\ref{Eq:RX signal at k}), the following correspondence can be established
\begin{eqnarray}
\left( \ma{A}, \ma{B}, \ma{\Omega}, \ma{C}^{\ma{A}}, \ma{C}^{\ma{B}}\right) \leftrightarrow (\ma{HS}, \ma{X}, \ma{G}, \ma{P}, \ma{W}).
\label{EQ:PARATUCK_correspondece}
\end{eqnarray}

To jointly estimate the involved channels and the transmitted data symbol matrices, we formulate the following optimization problem:
\begin{equation}
(\hat{\overline{\ma{H}}},\hat{\ma{G}},\hat{\ma{X})} =\underset{\overline{\ma{H}},\ma{G},\ma{X}}{\textrm{min}} \,\,\sum\limits_{k=1}^K\Big\| \ma{Y}_k  -  \overline{\ma{H}}D_k(\ma{S})\ma{G}D_k(\ma{W})\ma{X}^{\text{T}} \Big\|^2_F.
   \label{Eq: cost_func PT2}
\end{equation}
Since the problem is nonlinear with respect to the unknown matrices %(channel matrices and symbol matrix), 
we exploit its multilinear structure and recast it as a set of three conditionally linear problems solved iteratively \cite{Pierre_2009}. Specifically, one matrix is updated at each iteration by fixing the other two at the values obtained in previous steps. 
\subsection{Estimation of the RIS-Rx channel $\overline{\ma{H}}$}

Consider the received signal in (\ref{Eq:RX signal at k}). After $K$ blocks, we can collect all signals in the 1-mode unfolding as  
\begin{equation}
%\begin{split}
[\ten{Y}]_{(1)} \doteq \big[\ma{Y}_1, \ldots, \ma{Y}_K\big]
=\overline{\ma{H}}\ma{F}^{\text{T}}\, \, \in \mathbb{C}^{M_R \times TK},
\label{Eq: Estimate_H}  
%\end{split}
\end{equation}
where 
% \begin{equation}
%     \ma{F} \doteq \left[
%     \begin{array}{c}
%        \ma{X}D_1(\ma{W})\ma{G}^{\text{T}}D_1(\ma{P})\\
%        \vdots \\
%        \ma{X}D_K(\ma{W})\ma{G}^{\text{T}}D_K(\ma{P})
%     \end{array}
%     \right] \;\; \in \; \mathbb{C}^{TK \times N}.\label{eq:Fmatrix}
% \end{equation}
\begin{equation}
    \Big[\left(\ma{X}D_1(\ma{W})\ma{G}^{\text{T}}D_1(\ma{P})\right)^\text{T} \; \dots \; \left(\ma{X}D_K(\ma{W})\ma{G}^{\text{T}}D_K(\ma{P})\right)^\text{T}\Big]^\text{T}
    \label{eq:Fmatrix}
\end{equation}
 To estimate $\overline{\ma{H}}$ we need to solve the following LS problem
\begin{equation}
\Hat{\overline{\ma{H}}} = \underset{\overline{\ma{H}}}{\arg\min} \,\, \left\|[\ten{Y}]_{(1)} -\overline{\ma{H}}\ma{F}^{\text{T}}\right\|_F^2,
\label{Eq:func costH_PT2}
\end{equation}
the solution of which is given by
\begin{equation}
    \Hat{\overline{\ma{H}}} = [\ten{Y}]_{(1)}\left( \ma{F}^{\text{T}}\right)^{\dagger}.
    \label{Eq: Estimate_HBar_LS}  
\end{equation}
From $\Hat{\overline{\ma{H}}}$ the estimate of $\Hat{\ma{H}}$ is given filtering the block diagonal phase shift $\ma{S}$ as $\Hat{\ma{H}} = \Hat{\overline{\ma{H}}}\ma{S}^{\text{H}}\;.$
% \begin{equation}
%     \Hat{\ma{H}} = \Hat{\overline{\ma{H}}}\ma{S}^{\text{H}}\; .
%     \label{Eq: Estimates_H_LS}
% \end{equation}

\subsection{Symbol estimation}
In a similar way, we get the $2$-mode unfolding as follows
\begin{equation}
%\begin{split}
[\ten{Y}]_{(2)} \doteq \big[\ma{Y}_1^\text{T}, \ldots, \ma{Y}_K^\text{T}\big]
 =\ma{X}\ma{E}^{\text{T}} \, \, \in \mathbb{C}^{T \times M_RK},
\label{Estimate_X}
%\end{split}
\end{equation}
where
% \begin{equation}
% \ma{E} \doteq \left[\begin{array}{c}
% \overline{\ma{H}}D_1(\ma{S})\ma{G}D_1(\ma{W})\\
% %\ma{H}D_2(\ma{S})\ma{G}D_2(\ma{W})\\
% \vdots\\
% \overline{\ma{H}}D_K(\ma{S})\ma{G}D_K(\ma{W})
% \end{array}\right]\in \mathbb{C}^{M_RK \times M_T}.
% \label{eq:Ematrix}
% \end{equation}
\begin{equation}
\Big[\left(\overline{\ma{H}}D_1(\ma{S})\ma{G}D_1(\ma{W})\right)^\text{T} \; \dots \; \left(\overline{\ma{H}}D_K(\ma{S})\ma{G}D_K(\ma{W})\right)^\text{T}\Big]^\text{T}
    \label{eq:Ematrix}
\end{equation}
To estimate $\ma{X}$, we solve the following LS problem
\begin{equation}
\Hat{\ma{X}} = \underset{\ma{X}}{\arg\min} \,\, \left\|[\ten{Y}]_{(2)} -\ma{X}\ma{E}^{\text{T}}\right\|_F^2,
\label{func costX_PT2}
\end{equation}
the solution of which is given by: 
\begin{equation}
    \Hat{\ma{X}} = [\ten{Y}]_{(2)}\left( \ma{E}^{\text{T}}\right)^{\dagger}.
        \label{X_hat}
\end{equation}

\subsection{Estimation of the Tx-RIS channel $\ma{G}$}
To estimate the Tx-RIS channel matrix $\ma{G}$, let us apply the $\textrm{vec}(.)$ and the properties (\ref{Eq:Propertie Vec General}) and (\ref{Eq:Propertie Vec restrict}), such that the vectorized form of the received signal (\ref{Eq:RX signal at k}), is given as
\begin{equation}
\begin{aligned}
\textrm{vec}\left(\ma{Y}_k\right)  & = (\ma{X} \otimes \overline{\ma{H}})\textrm{vec}\left(D_k(\ma{P})\ma{G}D_k(\ma{W})\right)\\
& = (\ma{X} \otimes \overline{\ma{H}})\left(D_k(\ma{W}) \otimes D_k(\ma{P})\right)\textrm{vec}(\textbf{G})\, ,
\end{aligned}
\label{VEC_Equation Key 35}
\end{equation}
Note that $\left(D_k(\ma{W}) \otimes D_k(\ma{P})\right)$ is actually $\textrm{diag}\left(\ma{W}_{k\cdot}^{\text{T}} \otimes \ma{P}_{k\cdot}^{\text{T}}\right)$. Then, applying the property (\ref{Eq:Propertie diag(a)b}) to  (\ref{VEC_Equation Key 35}) yields
\begin{equation}
\textrm{vec}\left(\ma{Y}_k\right) = (\ma{X} \otimes \overline{\ma{H}})\textrm{diag}(\textrm{vec}(\ma{G}))\big(\ma{W}_{k\cdot}^{\text{T}} \otimes \ma{P}_{k\cdot}^{\text{T}}\big), 
\label{Concatenating VEC(35)}
\end{equation}
By collecting the vectorized version of the $K$ received signal blocks, we can build the 3-mode unfolding of the received signal tensor, which can be expressed as
\begin{equation}
\begin{aligned}
[\ten{Y}]_{(3)} &\doteq \big[\textrm{vec}\left(\ma{Y}_1\right), \ldots, \textrm{vec}\left(\ma{Y}_K\right)\big]\\
&= (\ma{X} \otimes \overline{\ma{H}})\textrm{diag}(\textrm{vec}(\ma{G}))\ma{\Psi} \, ,\, \in \,  \mathbb{C}^{TM \times K}
\end{aligned}
\label{VEC to ESTIMATE H}
\end{equation}
where 
\begin{equation}
\begin{aligned}
\ma{\Psi} & \doteq \left[\ma{W}_{1\cdot}^{\text{T}} \otimes \ma{P}_{1\cdot}^{\text{T}}, \ldots, \ma{W}_{K\cdot}^{\text{T}} \otimes \ma{P}_{K\cdot}^{\text{T}}\right]\\
& =\ma{W}^{\text{T}} \diamond \ma{P}^{\text{T}} \in \mathbb{C}^{M_TN \times K}. \label{eq:defPsi}
\end{aligned}
\end{equation}
Then, vectorizing (\ref{VEC to ESTIMATE H}) yields
\begin{equation}
\begin{aligned}
\textrm{vec}\left([\ten{Y}]_{(3)}\right) &= \big[\ma{\Psi}^\text{T} \diamond (\ma{X} \otimes \overline{\ma{H}})\big]\textrm{vec}(\ma{G}) + \textrm{vec}\left(\ma{B}_3\right).%\\ 
%& = \left[\ma{\Psi}^\textrm{T} \diamond (\ma{X} \otimes \ma{H})\right]\textrm{vec}(\ma{G}) 
\end{aligned}
\label{Estimate_H}
\end{equation}
Finally, an estimate of $\ma{G}$ in the LS sense can be obtained by solving the following problem
%{\setlength{\mathindent}{0cm}
\begin{equation}
\Hat{\ma{G}} = \underset{\ma{G}}{\arg\min} \,\, \left\|\textrm{vec}\left([\ten{Y}]_{(3)}\right) -\big[\ma{\Psi}^\text{T} \diamond (\ma{X} \otimes \overline{\ma{H}})\big]^{\dagger} \textrm{vec}(\ma{G})\right\|_F^2,
\label{func costG_PT2}
\end{equation}
%}
the solution of which is given by
\begin{equation}
    %\textrm{vec}(\ma{G}) = \ma{C}^{\dagger} \textrm{vec}(\ma{Y}_3).
     \Hat{\ma{G}} =\textrm{unvec}_{N \times L}\Big(\left[\ma{\Psi}^\text{T} \diamond (\ma{X} \otimes \overline{\ma{H}})\right]^{\dagger} \textrm{vec}([\ten{Y}]_{(3)})\Big).
    \label{Eq: Estimate G_hat}
\end{equation}
{\color{black}
The proposed semi-blind receiver utilizes equations (\ref{Eq: Estimate_HBar_LS}), (\ref{X_hat}), and (\ref{Eq: Estimate G_hat}) to estimate channel matrices $\ma{H}$ and $\ma{X}$, as well as the symbol $\ma{G}$ through a trilinear alternating least squares based estimation scheme, which is known as the \ac{TALS} receiver summarized in the algorithm in Algorithm \ref{PsCode_PARATUCK}. %This algorithm involves a three-step estimation procedure that estimates one matrix at each step while keeping the other two matrices constant to the values obtained at the previous estimation steps. It is worth noting that the proposed \ac{TALS} receiver is a semi-blind method that does not require any training sequences. You can find a summary of the receiver algorithm in Algorithm \ref{PsCode_PARATUCK}.
The stopping criterion used in this work is based on the normalized squared error measure, which is computed at the end of each iteration. The error measure is denoted as $\epsilon_{(i)}$ and is given by the following equation: $\epsilon_{(i)}=\sum\limits_{k=1}^K\|\ma{Y}_k - \Hat{\ma{Y}}_{k_{(i)}}\|_F^2/\|\ma{Y}_k\|_F^2$.
Here, $\Hat{\ma{Y}}_{k_{(i)}} = \hat{\ma{H}}_{(i)}D_k(\ma{S})\hat{\ma{G}}_{(i)}D_k(\ma{W})\hat{\ma{X}}^{\textrm{T}}_{(i)}$. Convergence is declared when the difference between the reconstruction errors of two successive iterations falls below a threshold, which is represented by $| \epsilon_{(i)} - \epsilon_{(i-1)}| \leq \delta$. In this work, we have assumed $\delta = 10^{-5}$.
}

\IncMargin{1em}
\begin{algorithm}[!t]
\footnotesize{
	\DontPrintSemicolon
	\SetKwData{Left}{left}\SetKwData{This}{this}\SetKwData{Up}{up}
	\SetKwFunction{Union}{Union}\SetKwFunction{FindCompress}{FindCompress}
	\SetKwInOut{Input}{input}\SetKwInOut{Output}{output}
	\textbf{Procedure}\\
	\Input{$i = 0$; \textrm{Initialize} $\Hat{\ma{G}}_{(i=0)}$ \textit{and} $\Hat{\ma{X}}_{(i=0)}$}
	\Output{$\Hat{\ma{H}}$, $\Hat{\ma{G}}$ and $\Hat{\ma{X}}$}
	\BlankLine

	\Begin{
		$i = 1 ;$\;
		\While{$\|e(i) - e(i-1)\| \geq \delta$}{
			\vspace{2ex}
			\begin{enumerate}
				\item [1.] \textrm{Using} $\Hat{\ma{G}}_{(i-1)}$ \textrm{and} $\Hat{\ma{X}}_{(i-1)},$ \textrm{compute}\\ $\Hat{\ma{F}}_{(i-1)}$ \textrm{from} (\ref{eq:Fmatrix}) \textrm{and find \\ a least squares estimate of} $\ma{H}$:\\
				\vspace{1ex}
				$\Hat{\overline{\ma{H}}}_{(i)} = \ma{Y}_1\left( \Hat{\ma{F}}^{\textrm{T}}_{(i-1)}\right)^{\dagger} \rightarrow$
                    $\Hat{\ma{H}} = \Hat{\overline{\ma{H}}}\ma{S}^{\text{H}}$
				\vspace{2ex}
				\item[2. ] \textrm{Using} $\Hat{\ma{H}}_{(i)}$ and $\Hat{\ma{X}}_{(i-1)}$, find\\ 
				\textrm{a least squares estimate of $\ma{G}$}:\\
				\vspace{1ex}
				$\textrm{vec}(\Hat{\ma{G}}_{(i)})= \left[\ma{\Psi}^\textrm{T} \diamond (\ma{X}_{(i-1)} \otimes \ma{H}_{(i)})\right]^{\dagger} \textrm{vec}(\ma{Y}_3)$%\\
			\vspace{2ex}
				\item[3.] \textrm{Using} $\Hat{\ma{G}}_{(i)}$ and $\Hat{\ma{H}}_{(i)},$ \textrm{compute}\\ $\Hat{\ma{E}}_{(i)}$ \textrm{from} (\ref{eq:Ematrix}) \textrm{and find \\ a least squares estimate of} $\ma{X}$:\\
				\vspace{1ex}
				$\Hat{\ma{X}}_{(i)} = \ma{Y}_2\left( \Hat{\ma{E}}^{\textrm{T}}_{(i)}\right)^{\dagger}$
				\vspace{2ex}
				\item[4:] $i \leftarrow i+1$
				\item[5:]\textrm{Repeat steps} $1$ to $4$ \textrm{until convergence.}
			\end{enumerate}
			\textbf{end}
		}
	\textbf{end}	
	}
	\caption{TALS}\label{PsCode_PARATUCK}
	\label{Algorithm:TALS}
	}
\end{algorithm}\DecMargin{0.8em}
{\color{black}\subsection{Identifiability and complexity }
The joint recovery of $\ma{H}$, $\ma{X}$, and $\ma{G}$ requires that the three LS problems in (\ref{Eq:func costH_PT2}), (\ref{func costX_PT2}) and (\ref{func costG_PT2})  have unique solutions, respectively. More specifically, the uniqueness of $\ma{H}$ requires that $\ma{F}$ defined in (\ref{eq:Fmatrix}) have a full column rank, which implies $TK \geq N$, while the uniqueness of $\ma{G}$ requires that $\left[\ma{\Psi}^\textrm{T} \diamond (\ma{X} \otimes \ma{H})\right]$ have a full column rank, which implies $TKM_R \geq M_TN$. Likewise, the uniqueness of $\ma{X}$ requires that $\ma{E}$ defined in (\ref{eq:Ematrix}) be of full column-rank, which implies $M_RK \geq M_T$. %Note that the number $K$ of transmitted blocks is the common parameter in these conditions, which must be satisfied simultaneously.
In summary, the following conditions must hold for the joint uniqueness of $\ma{H}$, $\ma{G}$, and $\ma{X}$: $TK \geq N, \quad TKM_R \geq M_TN, \quad M_RK \geq M_T.$
% \begin{equation}
% TK \geq N, \quad TKM_R \geq M_TN, \quad M_RK \geq M_T.
% \label{EQ:indenti Condition}
% \end{equation}
These conditions establish useful trade-offs involving the time diversities (parameters $K$ and $T$) and spatial diversities (parameters $N$, $M_R$, $M_T$) for the joint recovery of the channel and symbol matrices.% More specifically, reducing the number of blocks $K$ and/or the number of symbol periods $T$ can be compensated by a corresponding increase in the number of \ac{BS} antennas $M_R$. 
%Under these conditions stated previously, the estimates of $\ma{G}$, $\ma{H}$, and $\ma{X}$ provided by Algorithm \ref{Algorithm:TALS} are affected by scaling ambiguities that compensate each other, as follows
% \begin{equation}
% \Hat{\ma{H}}\boldsymbol{\Delta}_{H} = \ma{H}, \quad
% \Hat{\ma{X}}\boldsymbol{\Delta}_{X} = \ma{X}, \quad
% \boldsymbol{\Delta}_{H}^{-1}\Hat{\ma{G}}\boldsymbol{\Delta}_{X}^{-1} = \ma{G},\label{eq:ambiguities_paratuck}
% \end{equation} 
%where $\boldsymbol{\Delta}_{G}$, $\boldsymbol{\Delta}_{H}$, and $\boldsymbol{\Delta}_{X}$ are diagonal matrices. The scaling ambiguity can be addressed by assuming that the first row of the symbol matrix \(\ma{X}_{1.} \in \mathbb{C}^{1 \times L}\) contains identification symbols that are known to the receiver.% It is important to note that the columns of \(\ma{X} \in \mathbb{C}^{T \times M_T}\) correspond to the \(M_T\) data streams that are spatially multiplexed at the transmitter. A straightforward approach is to set \(\ma{X}_{1.} = [1, 1, \ldots, 1]\). This allows us to determine \(\boldsymbol{\Delta}_{X}\) from the first row of the estimated symbol matrix \(\Hat{\ma{X}}\) once Algorithm \ref{Algorithm:TALS} has converged. We can then cancel out the scaling factor by applying normalization.
The complexity of the algorithm \ref{Algorithm:TALS} is dominated by the pseudo-inverse calculation in step $2$. Considering that the
computational complexity  of the pseudo-inverse is equivalent to that of the singular value
decomposition (SVD), the complexity of our proposed solution is $\mathcal{O}(TKM_R(M_TN)^2)$}.

\section{Performance evaluation}
\label{SEC:Results}
We evaluate the performance of the proposed semi-blind receivers. The \ac{CE} accuracy is evaluated in terms of the normalized mean square error (NMSE) given by $\textrm{NMSE}(\Hat{\ma{\Pi}}) = \frac{1}{R}\sum_{r=1}^{R} \dfrac{\|\ma{\Pi}^{(r)} - \Hat{\ma{\Pi}}^{(r)}\|_F^2}{ \|\ma{\Pi}^{(r)}\|_F^2}$,
% \begin{equation}
%     \textrm{NMSE}(\Hat{\ma{\Pi}}) = \frac{1}{R}\sum_{r=1}^{R} \dfrac{\|\ma{\Pi}^{(r)} - \Hat{\ma{\Pi}}^{(r)}\|_F^2}{ \|\ma{\Pi}^{(r)}\|_F^2}\, ,
% \end{equation}
where $\ma{\Pi} = \ma{H}, \ma{G}$ and $\Hat{\ma{\Pi}}^{(r)}$ denotes the estimation of the channels at the $r$th run, and $R$ denotes the number of Monte-Carlo runs.  We also evaluate both the \ac{SER}  performance and complexity in terms of the number of iterations as a function of the signal-to-noise ratio (SNR). Regarding the channel model, we consider the Rayleigh fading case (i.e. the entries of channel matrices are independent and identically distributed zero-mean circularly-symmetric complex Gaussian random variables. All results represent an average of at least $R=3000$ Monte Carlo runs. Each run corresponds to an independent realization of the channel matrices, transmitted symbols, and the noise term.

\begin{figure}
    \centering
    \includegraphics[scale=0.45]{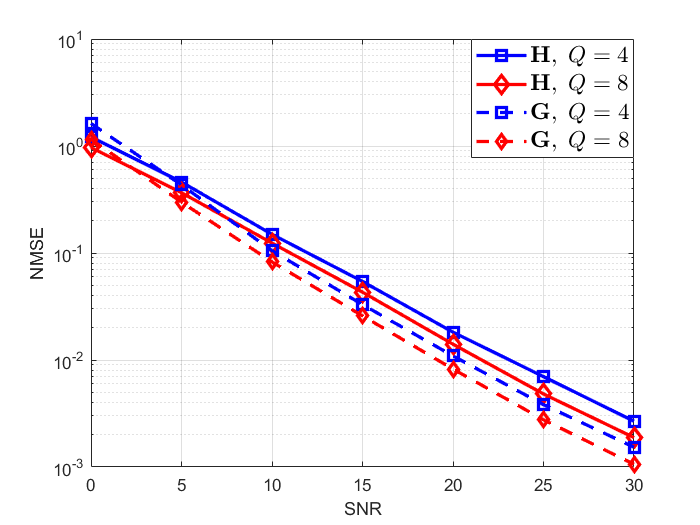}
    \caption{NMSE vs. SNR}
    \label{Fig: H and G}
\end{figure}

\begin{figure}
    \centering
    \includegraphics[scale=0.45]{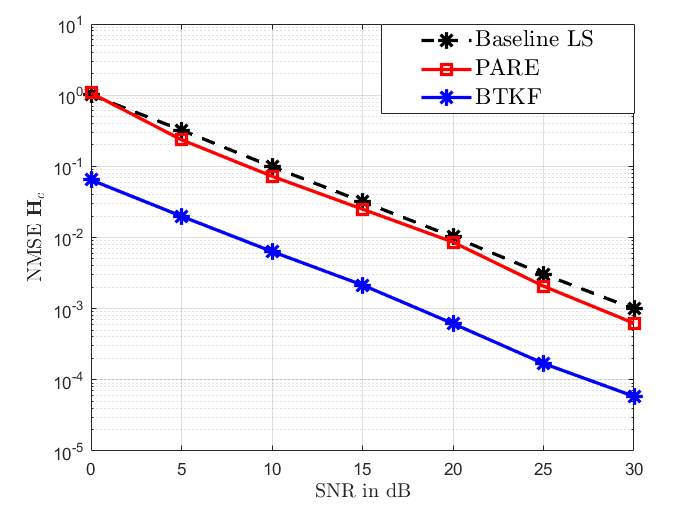}
    \caption{NMSE of equivalent channel vs. SNR}
    \label{Fig: theta}
\end{figure}

We assume $M_R=5$, $M_T=2$, $K=20$, $T=10$, and $N=16$ with $Q={4,8}$. {\color{black} The scattering matrix for the group $q$, $\ma{S}_q$, as well as the rotation and coding matrices $\ma{P}$ and $\ma{W}$, respectively, can be a proper DFT.} Figure \ref{Fig: H and G} depicts the channel estimation performance in terms of \ac{NMSE}. The accuracy of channel estimation $\ma{G}$ is better than that of $\ma{H}$, as the former requires a spatial filter step to obtain this channel estimation. 
{\color{black}Considering the composed channel \(\ma{H}_c = \ma{G}^{\text{T}} \otimes \ma{H}\), we present a comparison of our proposed method with existing approaches in the literature, specifically the least squares (LS) method \cite{B_Clerckc_CE_TSP_24} and the block Tucker Khatri-Rao factorization (BTKF) method \cite{Sokal_BD_RIS_2024}, as shown in Figure \ref{Fig: theta}. We assume the parameter set \(Q = 2\), \(N = 16\), \(M_R = M_T = 4\), and \(KT = M_T N_q^2 Q\) to ensure a fair comparison with the reference methods. First, we can see that the proposed method outperforms the LS method. This improvement is due to the noise rejection gain achieved by our iterative approach, which enables decoupled channel estimation. We can also note that the proposed PARE receiver exhibit has a performance gap in comparison to the reference BTKF channel estimation method. However, it is important to note that both LS and BTKF are pilot-assisted methods, while PARE is a semi-blind receiver that uses actual data symbols to estimate the individual channels. This means that, compared to BTKF, the proposed receiver offers a higher spectral efficiency by allowing joint symbol decoding and channel estimation. More specifically, the PARE receiver enables decoding of \( M_{T} \) useful data streams during the channel estimation process, therefore exhibiting a smaller data decoding delay compared to competing pilot-assisted methods for BD-RIS. Additionally, PARE can operate under more flexible system parameter settings than its competitors. Although we have selected \(KT = M_T N_q^2 Q\) $(K = 64 \; \text{and} \; T = 8)$ to meet the constraints of the competing methods, PARE can operate under different choices of \(K\) and \(T\) allowing to trade performance and spectral efficiency. %However, in the paper, we choose \(K\) to create an orthogonal design for \(\ma{\Psi}\).
}
%{\color{black}Considering the composed channel $\ma{H}_c = \ma{G}^{\text{T}} \otimes \ma{H}$, we plot in Figure \ref{Fig: theta} the performance of our proposed method and compare it with existing approaches in the literature, the LS method \cite{B_Clerckc_CE_TSP_24} and BTKF \cite{Sokal_BD_RIS_2024}. Using the parameters $Q = 2$, $N = 16$, $M_R = M_T = 4$, and $KT = M_TN_q^2Q$, due to the time restriction required by the benchmark methods, we can observe that our proposed method outperformed the LS method. This occurs since our proposed solution enjoys the noise rejection gain provided by the iterative method that enables decoupled channel estimation. The BTKF method surpasses our approach, which can be attributed to its exploitation of the block Kronecker structure of the composed channel. However, while both LS and BTKF are pilot assisted methods our solution is a semi-blind approach. Furthermore, system parameters choice our method it more flexible than competitors such that the $KT = M_TN_q^2Q$ was chosen to satisfy the restriction of the competitor methods.}  

%\vspace{-4ex}

\begin{figure}
    \centering
    \includegraphics[scale=0.45]{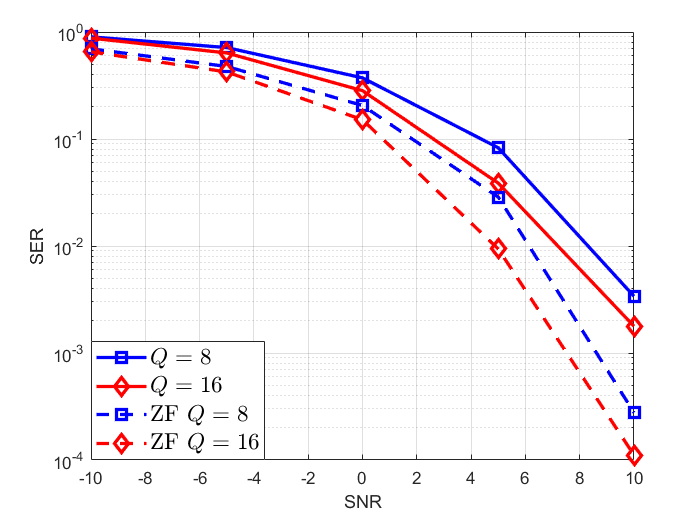}
    \caption{SER vs SNR}
    \label{Fig: SER}
\end{figure}

\begin{figure}
    \centering
    \includegraphics[scale=0.45]{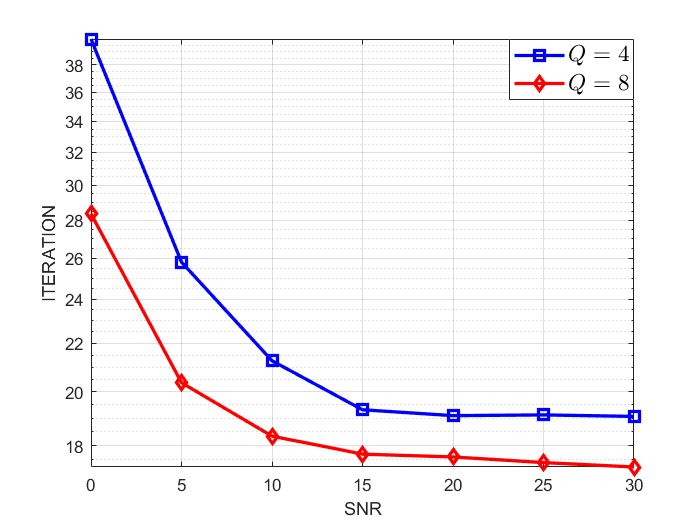}
    \caption{Average runtime vs. SNR}
    \label{Fig: IT}
\end{figure}

Figure \ref{Fig: SER} depicts the SER performance for two different numbers of BD-RIS groups. The performance of the zero-forcing (ZF) receiver under perfect channel knowledge is also shown as a reference. Note that increasing the number of groups produces improved SER and implies a reduced complexity in the number of iterations, as illustrated in the results of Figure~\ref{Fig: IT}. From this figure, we can also see that the convergence of Algorithm 1 is usually achieved within 20 iterations in the moderate SNR range for the considered parameter settings.

%\vspace{-4ex}
%\vspace{-4ex}

\section{Conclusion}
{\color{black} This paper addressed semi-blind channel estimation for BD-RIS. The proposed PARE receiver provides joint channel and symbol estimation based on an alternating estimation scheme derived from a PARATUCK tensor modeling of the received signals. Our proposed receiver enjoys more flexible system parameter choices than traditional LS-based estimation while allowing a data-aided channel estimation without a prior training stage. A perspective of this work includes alternative BD-RIS design structures and extensions to multi-user scenarios.
}
% \section*{Acknowledgment}

% The authors acknowledge the partial support of Fundação Cearense de Apoio ao Desenvolvimento Científico e Tecnológico (FUNCAP) under grants FC3-00198-00056.01.00/22 and  ITR-0214-00041.01.00/23, and the National Institute of Science and Technology (INCT-Signals) sponsored by Brazil's National Council for Scientific and Technological Development (CNPq) under grant 406517/2022-3. This work is also partially supported by CNPq under grants 312491/2020-4 and 443272/2023-9.

\bibliographystyle{IEEEtran}
%\bibliography{IEEEexample}

\end{document}